\documentclass[onecolumn,preprint,a4paper,12pt,aps]{revtex4}
\usepackage[left=19mm,right=19mm,top=17mm,bottom=19mm,includehead,includefoot]{geometry}

\usepackage[dvips]{graphics,color}
\usepackage{longtable}
\usepackage{enumerate}
\usepackage{amsmath}

\usepackage{here}

\usepackage{graphicx}
\usepackage{dcolumn}
\usepackage{bm}

\begin{document}

\title{Local Relativistic Exact Decoupling}

\author{Daoling Peng$^{*}$}

\author{Markus Reiher\footnote{Corresponding authors, email:
daoling.peng@phys.chem.ethz.ch, markus.reiher@phys.chem.ethz.ch}}

\affiliation{ETH Zurich, Laboratorium f\"{u}r Physikalische Chemie,
Wolfgang-Pauli-Strasse 10, CH-8093 Zurich, Switzerland\\}

\begin{abstract}


{\small
We present a systematic hierarchy of approximations for {\it local} exact-decoupling of four-component 
quantum chemical Hamiltonians based
on the Dirac equation. Our ansatz reaches beyond the trivial local approximation that is based
on a unitary transformation of only the atomic block-diagonal part of the Hamiltonian. 
Systematically, off-diagonal Hamiltonian matrix blocks can be subjected to a unitary transformation
to yield relativistically corrected matrix elements. The full hierarchy is investigated with
respect to the accuracy reached for the electronic energy and for selected molecular properties on a balanced
test molecule set that comprises molecules with heavy elements in different bonding situations.
Our atomic (local) assembly of the unitary exact-decoupling transformation --- called
local approximation to the unitary decoupling transformation (DLU) --- provides an excellent local approximation for any relativistic
exact-decoupling approach. Its order-$N^2$ scaling can be further reduced to linear
scaling by employing a neighboring-atomic-blocks approximation.
Therefore, DLU is an efficient relativistic method well suited for 
relativistic calculations on large molecules.
If a large molecule contains many light atoms (typically hydrogen atoms),
the computational costs can be further reduced by employing
a well-defined nonrelativistic approximation for these light atoms
without significant loss of accuracy.
We also demonstrate that the standard and straightforward transformation of only the atomic block-diagonal
entries in the Hamiltonian --- denoted DLH in this paper --- introduces an error that is on the order of the error of second-order Douglas--Kroll--Hess (i.e., DKH2)
when compared with exact-decoupling results. Hence, the local DLH approximation would be pointless in an exact-decoupling framework, but can
be efficiently employed in combination with the fast to evaluate DKH2 Hamiltonian in order to speed up calculations for which ultimate accuracy is not the major concern.
}


{\bf Date:} June 14, 2012 \\
{\bf Status:} printed in J.\ Chem.\ Phys.\ {\it 136} {\bf 2012} 244108

\end{abstract}

\maketitle

\section{Introduction}

The consideration of relativistic effects is essential to the proper understanding
of the chemistry of heavy-element containing molecules \cite{dyal07,reih09}.
The Dirac equation provides the relativistic quantum mechanical description of
a single electron in the presence of external electromagnetic potentials.
As the Dirac operator comprises four-dimensional operators,
its eigenfunctions possess four components.
The Coulomb interaction turned out to be sufficiently accurate for the
description of chemical phenomena and the corresponding many-electron Hamiltonian
is called the Dirac--Coulomb Hamiltonian.

A relativistic method based on the Dirac--Coulomb Hamiltonian 
is called a four-component method. It yields
negative-energy solutions, which are pathologic and of no use in chemistry.
Also, a large number of basis functions is required to properly describe
the negative-energy states.
These are two drawbacks of four-component methods that
motivated the development of two-component methods, which remove the negative energy solutions
and can, in principle, exactly reproduce the results of four-component calculations \cite{reih09}.

Several relativistic two-component methods were developed in the past decades.
One of the widely used approaches is the second-order
Douglas--Kroll--Hess method (DKH2) \cite{hess86_pra,hess89_pra}.
It employs the free-particle Foldy--Wouthuysen (FW) \cite{fowo50_pr} transformation as well as
sequential Douglas--Kroll \cite{dokr74_ap} transformations to decouple the four-component operator.
Higher-order \cite{naka00_cpl,naka00_jcp,wolf02_jcp,vanw04_jcp}
and even arbitrary-order \cite{reih04_jcp1,reih04_jcp2,wolf06_jcp1,wolf06_jcp2,reih07b,peng09_jcp}
DKH methods have been developed.
The zeroth-order regular approximation (ZORA) \cite{chan86,vanl93_jcp,vanl94_jcp}
is another highly successful relativistic two-component method. Within the ZORA framework, it is particularly
easy to implement the calculation of molecular properties; see Refs. \cite{auts11b,aqui11} for very recent examples and
Ref.\ \cite{auts07} for a review. 

In recent years it was shown that,
if the goal is to exactly decouple the four-component Hamiltonian matrix
instead of the Hamiltonian operator, the formulae of exact decoupling
can be directly obtained.
The Barysz--Sadlej--Snijders (BSS) \cite{bary97_ijqc,bary01_jms,bary02_jcp,bary07_cpl}
method aims at exact decoupling of the free-particle FW-transformed
four-component Hamiltonian matrix by a matrix operator of the form derived in Ref. \cite{heul86_jpb}.
The key matrix operator, which is used to construct the decoupling transformation matrix,
is obtained by solving an iterative equation.
However, invoking the free-particle FW transformation turns out to be
not necessary for the construction of the exact decoupling transformation
in a one-step protocol.
The pioneering work of this one-step transformation was provided by
Dyall \cite{dyal97_jcp,dyal98_jcp,dyal99_jcp,dyal01_jcp,dyal02_jcc}
in the form of the so-called normalized elimination of the small component (NESC) approach.
Later it was generalized to the so-called exact two-component (X2C) method
by several groups \cite{fila03_jcp,jens05,fila05_jcp,
kutz05_jcp,kutz06_mp,wliu06_jcp,fila07_tca,wliu07_jcp,saue07_jcp,peng07_jcp,wliu09_jcp,sikk09_jcp}.
Since the DKH method also employs matrix techniques to evaluate the Douglas--Kroll
transformation, it is also able to exactly decouple the four-component Hamiltonian
matrix. This has been shown within the arbitrary-order approach
\cite{reih04_jcp1,reih04_jcp2,wolf06_jcp1,wolf06_jcp2,mast07_jcp,peng09_jcp}.
For reviews of these developments see
Refs. \cite{hess00,reih06_tca,wliu10_mp,naka11_cr,reih11_wire,saue11_cpc,peng12_tca}.

Even within a scalar relativistic approximation,
where the spin-dependent operators are neglected,
the construction of the relativistic Hamiltonian matrix
dominates the step of the one-electron integral evaluation.
This is the case because the relativistic transformation is done within
an un-contracted basis set and involves many matrix manipulations
such as multiplication and diagonalization.
For a large molecule including only one or a few heavy atoms,
this procedure would waste computational resources
since the relativistic effects are highly local and only significant for heavy atoms.
It is therefore clear that a local relativistic method is desirable.
Even for a molecule containing only heavy atoms,
such a local relativistic approach would be preferable as we shall see.

The relativistic Hamiltonian operator is universal in coordinate space
and a local approach is unfeasible at the operator level.
However, since most quantum chemical calculations apply a linear combination of
atom-centered basis functions,
we may employ these basis functions to construct atomic projectors as
it is done in charge and spin population analysis \cite{herr05}.
For those Hamiltonian matrix blocks for which relativistic corrections are important,
we shall derive a local relativistic Hamiltonian matrix to evaluate it.
The locality is then exploited in the basis-function
space instead of coordinate space.
A crucial aspect to determine will be which block of the Hamiltonian matrix
requires a relativistic description and how to evaluate it.
Undoubtedly, the heavy atom diagonal blocks will require a relativistic description,
while the relativistic description of heavy atom off-diagonal blocks
depends on their contribution to physical observables
and on the feasibility of applying the relativistic description.
For relativistic transformation techniques, such as DKH and X2C,
it is impossible to apply the relativistic transformation to an off-diagonal block alone
without information about other blocks.
It requires the information of a full square matrix to construct the relativistic transformation.
Such a difficulty disappears
if one employs an operator-based relativistic method such as ZORA.
The off-diagonal block of the Hamiltonian matrix can be directly
computed from relativistic integrals.
However, the ZORA method provides a poor description of atomic inner-shell orbitals,
which carry the largest part of the relativistic effect.
In this sense, the exact decoupling methods are the best candidate for applying
a local relativistic scheme
since they do not neglect important relativistic effects.

All current exact decoupling methods employ the transformation technique.
It is therefore not trivial to apply the relativistic description
to heavy-atom off-diagonal Hamiltonian matrix elements.
However, in this work we propose that if the local approximation is applied to
the decoupling transformation instead of the Hamiltonian matrix,
one could obtain results which include the
relativistic description to heavy atom off-diagonal blocks.
We also note that the main effort for the construction of the relativistic Hamiltonian matrix
with exact-decoupling methods
is due to the evaluation of the decoupling transformation matrices.
Therefore, a local approximation to the decoupling transformation
would also lead to a significant reduction of the computational cost.

\section{Relativistically local structure of the Hamiltonian matrix}

The following notation will be used throughout this paper.
Upper case labels $A,B,C$ denote heavy atoms 
which require a relativistic description,
while lower case labels $a,b,c$ denote light atoms for which a (standard)
nonrelativistic description can be assumed to be accurate enough
based on numerical evidence compiled within computational chemistry in the past decades.
To be more precise, the former labels refer to nuclei with high nuclear charge numbers,
while the latter refer to those with low nuclear charge numbers (typically with nuclear
charges of less than two dozen protons). Of course, the distinction whether a nucleus is considered heavy or
light bears some sort of arbitrariness and it will be made on the basis of the accuracy required in a
quantum chemical calculation.
In general, there are then five different types of Hamiltonian matrix blocks
$H_{AA}$, $H_{AB}$, $H_{Aa}$, $H_{aa}$, and $H_{ab}$
that need to be considered.

The evaluation of the relativistic Hamiltonian matrix with exact-decoupling approaches
requires a set of square matrices.
They are the nonrelativistic kinetic energy matrix $T$,
\begin{equation}
T_{ij}=\frac{1}{2}\langle\lambda_i|\boldsymbol{p}^2|\lambda_j\rangle,
\end{equation}
the external potential matrix $V$, 
\begin{equation}
V_{ij}=\langle\lambda_i|\mathcal{V}|\lambda_j\rangle,
\end{equation}
and the overlap matrix $S$,
\begin{equation}
S_{ij}=\langle\lambda_i|\lambda_j\rangle,
\end{equation}
as well as an additional relativistic matrix $W$
\begin{eqnarray}
W_{ij}=\frac{1}{4c^2}\langle\lambda_i|
\boldsymbol{\sigma}\cdot\boldsymbol{p}
\mathcal{V}
\boldsymbol{\sigma}\cdot\boldsymbol{p}
|\lambda_j\rangle,
\end{eqnarray}
(note that all expressions are given in Hartree atomic units, in which the rest mass of the electron takes
a numerical value of one).
In the above equations, $\lambda_i$ denotes the $i$-th 2-spinor atom-centered basis function, $\mathcal{V}$ the external potential,
$c$ the speed of light, $\boldsymbol{\sigma}$ the vector of Pauli spin matrices. 
The basis functions $\lambda_i$ may be grouped according to the atomic nucleus to which they are assigned, which is key
to the local approaches discussed in the following.
The matrix representation of the relativistic Hamiltonian is then evaluated
as a function of the above-mentioned matrices
\begin{eqnarray}\label{r2ctrans}
H=H(T,V,W,S).
\end{eqnarray}


A straightforward idea for a trivial local approximation is to apply the evaluation of the relativistic
Hamiltonian matrix only to atomic (i.e., diagonal) blocks 
\begin{eqnarray}\label{r2ctrans_atom}
H_{AA}=H_{AA}(T_{AA},V_{AA},W_{AA},S_{AA}),
\end{eqnarray}
(instead of to the full matrix, but where $A$ may also represent a group of atoms)
and to ignore the relativistic correction to all off-diagonal blocks
\begin{eqnarray}
H_{AB}=T_{AB}+V_{AB}.
\end{eqnarray}
Adding more atoms to a group $A$ will improve the accuracy,
but the computational advances will be lost when a group $A$ becomes very large.
This approach was employed in Ref.\ \cite{pera04_jcp,pera05_jcp} and explored in detail in Ref.\ \cite{thar09_jcp}
for the low-order DKH method.
We refer to it as the {\it diagonal local approximation to the Hamiltonian} (DLH) matrix.
It is clear that the DLH approximation can also be applied to relativistic exact-decoupling approaches.
The DLH approximation works well at large interatomic distances, but
the difference to a reference energy increases with shorter distance.


However, relativistic corrections to off-diagonal blocks are also important
especially when inter-atomic distances are short.
The necessity of applying the relativistic description to off-diagonal blocks
has already been discussed in Ref.\ \cite{gagl98_cpl} with a so-called two-center approximation.
In Ref.\ \cite{gagl98_cpl}, the DKH transformation was applied to
pairs of atoms ($A\oplus B$)
\begin{eqnarray}\label{hsuperb}
H_{A\oplus B}&=&H(T_{A\oplus B},V_{A\oplus B},W_{A\oplus B},S_{A\oplus B}),
\end{eqnarray}
with
\begin{eqnarray}\label{hsuperb2}
H_{A\oplus B}&=&
\left(\begin{array}{cc} H_{AA} & H_{AB} \\
H_{BA} & H_{BB} \end{array}\right),
\end{eqnarray}
to obtain a relativistically corrected off-diagonal Hamiltonian matrix $H_{AB}$.
Unfortunately, such a two-center approximation introduces an inconsistent treatment
to diagonal blocks. For instance, $H_{AA}$ can be obtained from
either the pair ($A\oplus B$) or from another pair ($A\oplus C$).


So far, we have only discussed the $AA$ and $AB$ blocks which, by definition, require a relativistic
description for both atoms.
Since relativistic corrections for light atoms are very small and may be neglected,
$H_{aa}$ and $H_{ab}$ can be approximated by the nonrelativistic
Hamiltonian matrices
\begin{eqnarray}
H_{aa}&=&T_{aa}+V_{aa},\label{nrhaa}\\
H_{ab}&=&T_{ab}+V_{ab}.\label{nrhab}
\end{eqnarray}
A relativistic scheme for the heavy--light hybrid off-diagonal blocks $H_{Aa}$ should, however, be considered explicitly.


\section{Local Decomposition of the X-Operator}
The general expression of exact decoupling transformations
can be written as
\begin{eqnarray}\label{general_ur2c}
U=\left( \begin{array}{cc} Z^{+}\frac{1}{\sqrt{1+X^{\dag}X}} &
Z^{+}\frac{1}{\sqrt{1+X^{\dag}X}}X^{\dag} \\
-Z^{-}\frac{1}{\sqrt{1+XX^{\dag}}}X &
Z^{-}\frac{1}{\sqrt{1+XX^{\dag}}}\end{array}\right)\equiv
\left( \begin{array}{cc}
U^{LL} & U^{LS} \\
U^{SL} & U^{SS}
\end{array}\right).
\end{eqnarray}
$Z^{+}$ and $Z^{-}$ are two-component unitary operators,
and only $Z^{+}$ is required for the evaluation of the electrons-only Hamiltonian.
The $X$-operator generates the electronic small-component functions $\varphi^{+}_{S}$
from the large-component functions $\varphi^{+}_{L}$
\begin{eqnarray}
\varphi^{+}_{S}=X\varphi^{+}_{L}.
\end{eqnarray}
Two-dimensional Hamiltonians are then obtained by applying the
unitary transformation of Eq.\ (\ref{general_ur2c}) to blockdiagonalize ('$bd$') the four-dimensional Dirac-based Hamiltonian $D$
\begin{eqnarray}\label{transU}
H^{bd}=UDU^{\dag}.
\end{eqnarray}
It yields both the electrons-only Hamiltonian $H$ (as the upper left block in $H^{bd}$)
and the one for negative-energy solutions, although the latter will be discarded.

As discussed by Dyall in Ref. \cite{dyal98_jcp} for the NESC method,
the Breit-Pauli approximation to the $X$ matrix,
which is used to construct the relativistic transformation matrix,
is
\begin{eqnarray}\label{xeqiaprox}
X=I,
\end{eqnarray}
where $I$ denotes the identity matrix.
Such an approximation is not variationally stable
and thus cannot be used for variational calculations.
However, one could employ this approximation for light atoms only
\begin{eqnarray}
X_{aa}=I,
\end{eqnarray}
since the corrections to light atoms are small and may not affect
variational calculations.
This idea was suggested in Ref. \cite{peng07_jcp} for the X2C method
but no results were presented.


Although the lowest level approximation to $X$ Eq. (\ref{xeqiaprox})
did not provide useful results, the atomic approximation to the $X$
matrix gave results with very small errors.
As discussed by Dyall in Ref. \cite{dyal99_jcp,dyal01_jcp}, the $X$ matrix
can be approximated as the direct sum of atomic blocks
\begin{eqnarray}\label{localxmat}
X=X_{AA}\oplus X_{BB}\oplus\cdots
\end{eqnarray}
This approximation resembles the linear combination of atomic
four-spinors ansatz in four-component calculations.
The same approach was also employed in Ref.\ \cite{matv08_jcp}
in the sense of an atomic approximation to the projection on electronic states.
If the electronic states are further transformed to a two-component picture
by a renormalization matrix, this gives rise to a local X2C method.
As discussed in Refs.\ \cite{wliu06_jcp,peng07_jcp}, the local approximation
to the $X$ matrix Eq.\ (\ref{localxmat}) works well for
spectroscopic constants of diatomic molecules.

However, there are some drawbacks of the
local-$X$ matrix approximation to the X2C method.
Firstly, it only reduces the cost for the evaluation of $X$ matrix,
while the exact-decoupling transformation matrices are still of molecular dimension
as the renormalization matrix is not atomic block diagonal
even if the $X$ matrix is.
The computational demands for the evaluation of the relativistic approximation are thus still tremendous
for large molecules.
Secondly, the $X=I$ approximation did not provide a
satisfactory treatment of the $Aa$ blocks,
since it may suffer from variational collapse and
Eqs.\ (\ref{nrhaa}) and (\ref{nrhab}) cannot be fulfilled within this approximation.


\section{Local approximations to the exact-decoupling transformation}
In this article, we suggest an local approximation to the
exact-decoupling transformation.
We may approximate the unitary transformation by taking only the 'atomic' diagonal
blocks (all off-diagonal blocks are then set to zero)
\begin{eqnarray}\label{localU}
U=U_{AA}\oplus U_{BB}\oplus \cdots ~,
\end{eqnarray}
where the atomic unitary transformations $U_{AA}$ are obtained from the diagonalization of
the corresponding $D_{AA}$ blocks of matrix operator $D$.
We denote this approximation to $U$ as the {\it diagonal local approximation to the unitary decoupling transformation} (DLU).

If we now substitute the approximate unitary transformation of Eq.\ (\ref{localU})
into Eq.\ (\ref{transU}), the diagonal blocks, 
\begin{eqnarray}\label{exp_diag}
H^{bd}_{AA}=U_{AA}D_{AA}U_{AA}^{\dag},
\end{eqnarray}
turn out to be the same as in the DLH approach, while
the off-diagonal blocks then read
\begin{eqnarray}\label{exp_offdiag}
H^{bd}_{AB}=U_{AA}D_{AB}U_{BB}^{\dag},
\end{eqnarray}
which is to be compared with the expression 
\begin{eqnarray}\label{utranfull}
H^{bd}_{AB}=\sum_{I,J}U_{AI}D_{IJ}U_{JB}^{\dag},
\end{eqnarray}
where $U$ has not been approximated (here, $I$ and $J$ run over all atomic blocks).
Hence, the DLU approach also introduces a relativistic
description to off-diagonal 'interatomic' blocks when compared with the DLH approximation.

The cost for the assembly of the unitary transformation $U$ within the DLU approach
is then of order $N$, where $N$ measures the system size, namely the number of atoms.
It is linear scaling since the atomic approximation is directly
applied to the unitary transformation.
However, the next step of applying the unitary decoupling transformation
to obtain the relativistic Hamiltonian matrices is no longer linear scaling.
If no local approximation was applied for the unitary decoupling transformation,
according to Eq. (\ref{utranfull})
where the summation indices $I,J$ run over all atomic blocks,
the calculation of $H_{AB}$ matrix will require $2N^{2}$ matrix multiplications.
Since the number of matrices to be calculated is of order $N^{2}$,
the total cost of the relativistic transformation
without local approximation is therefore of order $N^{4}$.
If the DLU approximation is applied,
no summation will be needed and only two matrix multiplications will be required for each heavy--heavy block $H_{AB}$.
The total cost is then of order $N^{2}$.

If the distance of two atoms $A$ and $B$ is sufficiently large,
the relativistic description of $H_{AB}$ can be neglected.
Thus, we may define neighboring atomic pairs according to their distances.
Then, only the Hamiltonian matrix $H^{bd}_{AB}$ of neighboring pairs
requires the transformation
\begin{eqnarray}\label{dlunbtrans}
H^{bd}_{AB}=U_{AA}D_{AB}U_{BB}^{\dag},\ ~\forall A,B~\text{being neighbors},
\end{eqnarray}
whereas all other pairs are simply taken in their nonrelativistic form.
Since the number of neighboring pairs is usually
a linear function of system size.
The total cost is then reduced to order $N$.

To investigate the relativistic correction to hybrid $H_{Aa}$ blocks of the electrons-only Hamiltonian $H$,
we need to consider the explicit structure of the unitary decoupling transformation, 
\begin{eqnarray}\label{utrans}
H=
\left(\begin{array}{cc} U^{LL}, & U^{LS} \end{array}\right)
\left(\begin{array}{cc} ~V~ & ~T~ \\ T & (W\!-\! T) \end{array}\right)
\left(\begin{array}{c} U^{LL,\dag} \\ U^{LS,\dag} \end{array}\right) ,
\end{eqnarray}
where only the electronic part of the exact-decoupling transformation
represented by $U^{LL}$ and $U^{LS}$ [which are the upper part of Eq.\ (\ref{general_ur2c})]
is employed.
From Eq.\ (\ref{utrans}) it is easy to see that the electrons-only Hamiltonian reads
\begin{eqnarray}\label{eonlyham}
H&=&U^{LS}TU^{LL,\dag}+U^{LL}TU^{LS,\dag}-U^{LS}TU^{LS,\dag}\nonumber \\
&&+U^{LL}VU^{LL,\dag}+U^{LS}WU^{LS,\dag}.
\end{eqnarray}

In the nonrelativistic (NR) limit, where the speed of light approaches infinity,
we have
\begin{eqnarray}\label{nrlimit}
U^{LL,{\rm NR}}=I, ~~U^{LS,{\rm NR}}=I, ~~\mbox{and}~~ W^{\rm NR}=0.
\end{eqnarray}
If we insert them into the relativistic electrons-only Hamiltonian (\ref{eonlyham}),
we arrive at the nonrelativistic Hamiltonian matrix,
\begin{eqnarray}
H^{\rm NR}=T+V.
\end{eqnarray}
This suggests a solution for the relativistic treatment of the hybrid heavy--light blocks $H_{Aa}$,
because we may set all light-atom diagonal blocks in the unitary transformation to identity matrices
\begin{eqnarray}
U^{LL}_{aa}=I ~~\mbox{and}~~ U^{LS}_{aa}=I.
\end{eqnarray}
To reproduce the nonrelativistic Hamiltonian matrices of light-atom-only blocks,
we must set the $W$ matrix of light atoms to zero,
\begin{eqnarray}
W_{aa}=0.
\end{eqnarray}
The $W_{Aa}$ matrices are also set to zero
\begin{eqnarray}
W_{Aa}=0.
\end{eqnarray}
We denote this approximation for the treatment of hybrid blocks
as the DLU(NR) approach.
If we do not introduce any approximations to the $W$ matrices,
the transformation will yield a Hamiltonian that differs from the nonrelativistic Hamiltonian matrix
on the order of $1/c^{2}$, which corresponds to the order achieved in the Breit--Pauli approximation. We therefore
denote it as the DLU(BP) approach.

The final local approximation to the unitary matrices is then
\begin{eqnarray}
U^{LL}&=&U^{LL}_{AA}\oplus U^{LL}_{BB}\oplus\cdots U^{LL}_{aa}\oplus U^{LL}_{bb}\oplus\cdots ~, \label{localUL}\\
U^{LS}&=&U^{LS}_{AA}\oplus U^{LS}_{BB}\oplus\cdots U^{LS}_{aa}\oplus U^{LS}_{bb}\oplus\cdots ~, \label{localUS}
\end{eqnarray}
Where the atomic unitary transformation matrices of heavy atoms
are evaluated by the relativistic exact decoupling approaches.
The explicit expressions of Hamiltonian matrix blocks $H_{AA}$ and $H_{AB}$ are
\begin{eqnarray}
H_{AA}&=&U^{LS}_{AA}T_{AA}U^{LL,\dag}_{AA}+U^{LL}_{AA}T_{AA}U^{LS,\dag}_{AA} \nonumber \\
&&-U^{LS}_{AA}T_{AA}U^{LS,\dag}_{AA}+U^{LL}_{AA}V_{AA}U^{LL,\dag}_{AA} \nonumber \\
&&+U^{LS}_{AA}W_{AA}U^{LS,\dag}_{AA},\\
H_{AB}&=&U^{LS}_{AA}T_{AB}U^{LL,\dag}_{BB}+U^{LL}_{AA}T_{AB}U^{LS,\dag}_{BB} \nonumber \\
&&-U^{LS}_{AA}T_{AB}U^{LS,\dag}_{BB}+U^{LL}_{AA}V_{AB}U^{LL,\dag}_{BB} \nonumber \\
&&+U^{LS}_{AA}W_{AB}U^{LS,\dag}_{BB}.
\end{eqnarray}
If the DLU(NR) approach is employed, the unitary transformation matrices of light atoms
are simply set to identity matrices 
and the explicit form of $H_{Aa}$, $H_{aa}$, and $H_{ab}$ is
\begin{eqnarray}
H_{Aa}&=&U^{LL}_{AA}T_{Aa}+U^{LL}_{AA}V_{Aa},\\
H_{aa}&=&T_{aa}+V_{aa},\\
H_{ab}&=&T_{ab}+V_{ab}.
\end{eqnarray}
In the DLU(BP) approach, however, they read
\begin{eqnarray}
H_{Aa}&=&U^{LL}_{AA}T_{Aa}+U^{LL}_{AA}V_{Aa}+U^{LS}_{AA}W_{Aa},\\
H_{aa}&=&T_{aa}+V_{aa}+W_{aa},\\
H_{ab}&=&T_{ab}+V_{ab}+W_{ab}.
\end{eqnarray}

The relativistic transformation of a property operator $P$ \cite{baer90_jpb,wolf06_jcp1,wolf06_jcp2,mast07_jcp,reih08_cpl,peng09_jcp}
requires an additional matrix $Q$ and its matrix elements are given by
\begin{eqnarray}
Q_{ij}=\frac{1}{4c^2}\langle\lambda_i|
\boldsymbol{\sigma}\cdot\boldsymbol{p}
\mathcal{P}
\boldsymbol{\sigma}\cdot\boldsymbol{p}
|\lambda_j\rangle.
\end{eqnarray}
The relativistic corrected property matrix then reads
\begin{eqnarray}\label{proptrans}
P=U^{LL}PU^{LL,\dag}+U^{LS}QU^{LS,\dag}.
\end{eqnarray}

\section{DLU Evaluation within Exact-Decoupling Approaches}\label{sec_imple}

\subsection{The X2C approach}

The X2C method employs 
\begin{eqnarray}\label{u_x2c}
U_{\rm X2C}=
\left( \begin{array}{cc} \frac{1}{\sqrt{1+X^{\dag}X}} &
\frac{1}{\sqrt{1+X^{\dag}X}}X^{\dag} \\
-\frac{1}{\sqrt{1+XX^{\dag}}}X &
\frac{1}{\sqrt{1+XX^{\dag}}}\end{array}\right),
\end{eqnarray}
as the exact-decoupling unitary transformation.
The unitary operators for the electronic states are then
\begin{eqnarray}
U_{\rm X2C}^{LL}&=&\frac{1}{\sqrt{1+X^{\dag}X}}, \\
U_{\rm X2C}^{LS}&=&U^{LL}_{\rm X2C}X^{\dag}=\frac{1}{\sqrt{1+X^{\dag}X}}X^{\dag},
\end{eqnarray}
For an actual implementation, we need their matrix representations in a finite non-orthogonal space of
the atom-centered basis functions. The $X$-matrix is evaluated by diagonalizing the following
generalized eigenvalue equation
\begin{eqnarray}
\left(\begin{array}{cc}~V~&T\\ ~T~&(W\! -\! T)\end{array}\right)
\left(\begin{array}{c} C_{L}^{+} \\ C_{S}^{+}\end{array}\right)=
\left(\begin{array}{cc}S&0\\ 0&\frac{1}{2c^2}T\end{array}\right)
\left(\begin{array}{c} C_{L}^{+} \\ C_{S}^{+}\end{array}\right)\epsilon^{+},
\end{eqnarray}
where $C_{L}^{+}$ and $C_{S}^{+}$ denote the large- and small-component molecular-orbital
coefficients, respectively.
We obtain the $X$ matrix by the following equation
\begin{eqnarray}
\label{x2cstep}
X=C_{S}^{+}\left(C_{L}^{+}\right)^{-1}.
\end{eqnarray}
The matrix representation of $U^{LL}$ and $U^{LS}$ is then given by
\begin{eqnarray}
U_{\rm X2C}^{LL}&=&S^{1/2}(S^{-1/2}\widetilde{S}S^{-1/2})^{-1/2}S^{-1/2},\\
U_{\rm X2C}^{LS}&=&U_{\rm X2C}^{LL}X^{\dag},
\end{eqnarray}
with $\widetilde{S}$ defined as
\begin{eqnarray}
\widetilde{S}=S+\frac{1}{2c^2}X^{\dag}TX.
\end{eqnarray}
In our atomic approximation to the unitary transformation $U$, we need to evaluate the above equation
for each 'atomic' block of the original Hamiltonian in order to obtain
the diagonal blocks $U^{LL}_{AA}$ and $U^{LS}_{AA}$. The
decoupling transformations are then assembled according to Eqs.\ (\ref{localUL}) and (\ref{localUS})
and the relativistic Hamiltonian matrix follows from Eq.\ (\ref{utrans}).

\subsection{The BSS approach}

In the BSS approach, all matrices are transformed to an
orthonormal basis-function space first.
It is convenient to calculate the transformation matrix $K$ by
diagonalizing the nonrelativistic kinetic energy matrix
\begin{eqnarray}\label{orthonormalk}
TK=SKt,
\end{eqnarray}
since the eigenvalues $t$ can be used for later evaluation of
the free-particle FW transformation.
The eigenvector matrix $K$ has the following properties:
\begin{eqnarray}
K^{\dag}SK=I ~~\mbox{and}~~ K^{\dag}TK=t.
\end{eqnarray}
The free-particle FW (fpFW) transformation features four diagonal block matrices
\begin{eqnarray}
U_0=
\left( \begin{array}{cc}
U^{LL,{\rm fpFW}} & U^{LS,{\rm fpFW}} \\
U^{SL,{\rm fpFW}} & U^{SS,{\rm fpFW}} \end{array}\right)
=\left(\begin{array}{cc} \sqrt{\frac{E_0+c^2}{2E_0}}  &
\sqrt{\frac{E_0-c^2}{2E_0}}f \\ -\sqrt{\frac{E_0-c^2}{2E_0}} &
\sqrt{\frac{E_0+c^2}{2E_0}}f \end{array}\right),
\end{eqnarray}
with $E_0=\sqrt{2tc^2+c^4}$ and $f=\sqrt{2c^2/t}$.
It is then applied to yield a transformed four-dimensional Hamiltonian matrix $D_{0}$
\begin{eqnarray}\label{transd0}
D_{0}=U_{0}
\left(\begin{array}{cc}K&0\\ 0&K\end{array}\right)^{\dag}
\left(\begin{array}{cc}~V~&T\\ ~T~&(W\! -\! T)\end{array}\right)
\left(\begin{array}{cc}K&0\\ 0&K\end{array}\right)U_{0}^{\dag}.
\end{eqnarray}
Next, the exact-decoupling BSS transformation,
\begin{eqnarray}
U_{1}=
\left( \begin{array}{cc} \frac{1}{\sqrt{1+R^{\dag}R}} &
\frac{1}{\sqrt{1+R^{\dag}R}}R^{\dag} \\
-\frac{1}{\sqrt{1+RR^{\dag}}}R &
\frac{1}{\sqrt{1+RR^{\dag}}}\end{array}\right),
\end{eqnarray}
which has the same structure as the exact-decoupling transformation $U_{\rm X2C}$,
is applied.
The $R$ matrix is obtained by diagonalizing the free-particle FW
transformed four-component Hamiltonian matrix $D_{0}$
and employing a similar relation to the one in Eq.\ (\ref{x2cstep}).

After the exact-decoupling BSS transformation has been carried out, the Hamiltonian
matrix is back-transformed to the original (non-orthogonal) basis 
\begin{eqnarray}
H=
\left(\begin{array}{cc}K^{-1}&0\\ 0&K^{-1}\end{array}\right)^{\dag}
U_{1}D_{0}U_{1}^{\dag}
\left(\begin{array}{cc}K^{-1}&0\\ 0&K^{-1}\end{array}\right).
\end{eqnarray}
Consequently, we obtain the matrix form of $U_{\rm BSS}^{LL}$ and $U_{\rm BSS}^{LS}$
for the BSS approach as
\begin{eqnarray}
U_{\rm BSS}^{LL}&=&(K^{-1})^{\dag}(1+RR^{\dag})^{-\frac{1}{2}}
(U^{LL,{\rm fpFW}}+R^{\dag}U^{SL,{\rm fpFW}})K^{\dag},\\
U_{\rm BSS}^{LS}&=&(K^{-1})^{\dag}(1+RR^{\dag})^{-\frac{1}{2}}
(U^{LS,{\rm fpFW}}+R^{\dag}U^{SS,{\rm fpFW}})K^{\dag}.
\end{eqnarray}
The expressions for the relativistic Hamiltonian matrix (and property matrices)
are then as given above.

\subsection{The DKH approach}

The DKH approach features the same initial transformation as the BSS approach
to obtain the transformed Hamiltonian matrix $D_0$.
Subsequent decoupling transformations are expressed as
\begin{equation}
U^{(m)}=\prod_{k=m}^1 U_{k}
\end{equation}
with the generalized parametrization of the $U_k$ \cite{wolf02_jcp},
\begin{equation}
 U_k=\sum_{i=0}^{[m/k]}a_{k,i}\mathbf{W}_k^{i},
\end{equation}
where $\mathbf{W}_k$ are anti-hermitian matrix operators and $m$ is related to the order of the DKH expansion.
Since the expression of $\mathbf{W}_k$ is too lengthy to be presented here, we refer the reader to
Ref.\ \cite{peng09_jcp} for details.
If $m$ is large (strictly, if it approaches infinity), exact decoupling is achieved.
Usually, a low value for $m$ is often sufficient for calculations of relative energies and valence-shell properties
(i.e., $m$=1 for the original DKH2 approach).

With the complete decoupling DKH transformation written as
\begin{eqnarray}
U_{\rm DKH}=\left( \begin{array}{cc}
U^{LL}_{\rm DKH} & U^{LS}_{\rm DKH} \\
U^{SL}_{\rm DKH} & U^{SS}_{\rm DKH} \end{array}\right).
\end{eqnarray}
the matrix forms of $U_{\rm DKH}^{LL}$ and $U_{\rm DKH}^{LS}$ are
\begin{eqnarray}
U_{\rm DKH}^{LL}&=&(K^{-1})^{\dag}(U^{LL,(m)}U^{LL,{\rm fpFW}}
+U^{LS,(m)}U^{SL,{\rm fpFW}})K^{\dag},\\
U_{\rm DKH}^{LS}&=&(K^{-1})^{\dag}(U^{LL,(m)}U^{LS,{\rm fpFW}}
+U^{LS,(m)}U^{SS,{\rm fpFW}})K^{\dag}.
\end{eqnarray}

We should note that the finite-order DKH Hamiltonian matrices are not the result
of directly applying the decoupling transformation.
To give consistent results, the transformation in Eq.\ (\ref{utrans})
is only used to obtain the off-diagonal blocks
while the diagonal blocks are replaced by the traditional
finite-order DKH Hamiltonian.
However, for high-order calculations, the differences of with/without
replacing the diagonal blocks are very small.

\section{Results and discussion}
As we have shown, there exists a systematic hierarchy of several levels for the local relativistic 
approximation to Hamiltonian matrix elements. In this section, we study the accuracy of the different levels on a test molecule set
and introduce the following notation to distinguish the results:
\begin{enumerate}
\item {\bf FULL}: Full molecular relativistic transformation, no approximation.
\item {\bf DLU}: Diagonal local approximation to decoupling transformations,
$U^{LL}$ and $U^{LS}$ are evaluated for each atomic (diagonal) block. However, three variants are possible:
\begin{itemize}
\item[i)] {\bf DLU(ALL)}: Calculate relativistic transformation for all atoms
(all atoms are treated as 'heavy' elements). I.e.,
apply the 'atomic' unitary transformation to the diagonal and to all off-diagonal blocks of the Hamiltonian
\item[ii)] {\bf DLU(NB)}: Calculate relativistic transformation for all atoms
(all atoms are treated as 'heavy' elements), but apply the relativistic transformation only to blocks of the Hamiltonian assigned to neighboring atoms
(and to the diagonal blocks, of course) to achieve linear scaling.
\item[iii)] {\bf DLU($ABC$,BP)}: Distinguish heavy and light atoms,
where $ABC$ denote the heavy atoms and apply the BP approximation to the light atoms.
\item[iv)] {\bf DLU($ABC$,NR)}: Distinguish heavy and light atoms,
where $ABC$ denote the heavy atoms and apply the NR approximation to the light atoms.
\end{itemize}
\item {\bf DLH:} Diagonal local approximation to Hamiltonian matrix blocks.
The relativistic correction for off-diagonal 'interatomic' blocks are neglected.
\begin{itemize}
\item[i)] {\bf DLH(ALL) :} Calculate a relativistic description for all diagonal 'atomic' blocks.
\item[ii)] {\bf DLH($ABC$,NR/BP) :} Only heavy-atom blocks are considered for the relativistic description.
\end{itemize}
\end{enumerate}
Note that this classification system does not account for all possible local relativistic approximations:
i) $A$ could also represent a group of atoms and ii) additional
off-diagonal approximations may also be included.

We have implemented the local relativistic schemes discussed so far
for the X2C, BSS, and DKH exact-decoupling approaches
into the {\sc Molcas} programme package \cite{molcas3}, into which we have
recently implemented X2C, BSS, and the polynomial-cost DKH schemes \cite{peng12_tca}.
In our scalar-relativistic calculations the definition of the $W$ matrix was
\begin{eqnarray}
W_{ij}=\frac{1}{4c^2}\langle\lambda_i|
\boldsymbol{p}\cdot\mathcal{V}\boldsymbol{p}
|\lambda_j\rangle,
\end{eqnarray}
and $\lambda_i$ now refers to scalar basis functions instead of 2-spinor basis functions.
All other expressions derived in this work are the same for the scalar-relativistic X2C, BSS, and DKH variants.
We should mention that we have not considered the transformation of the two-electron integrals as it is
common in most exact-decoupling calculations. However, the local methods discussed here could be of value
also for one of the approximate methods available to transform the two-electron integrals (note that a full-fledged
transformation of the two-electron integrals would render the application of exact-decoupling methods inefficient and
a four-component approach should be employed instead).

\begin{figure}[H]
\scalebox{0.5}{\includegraphics{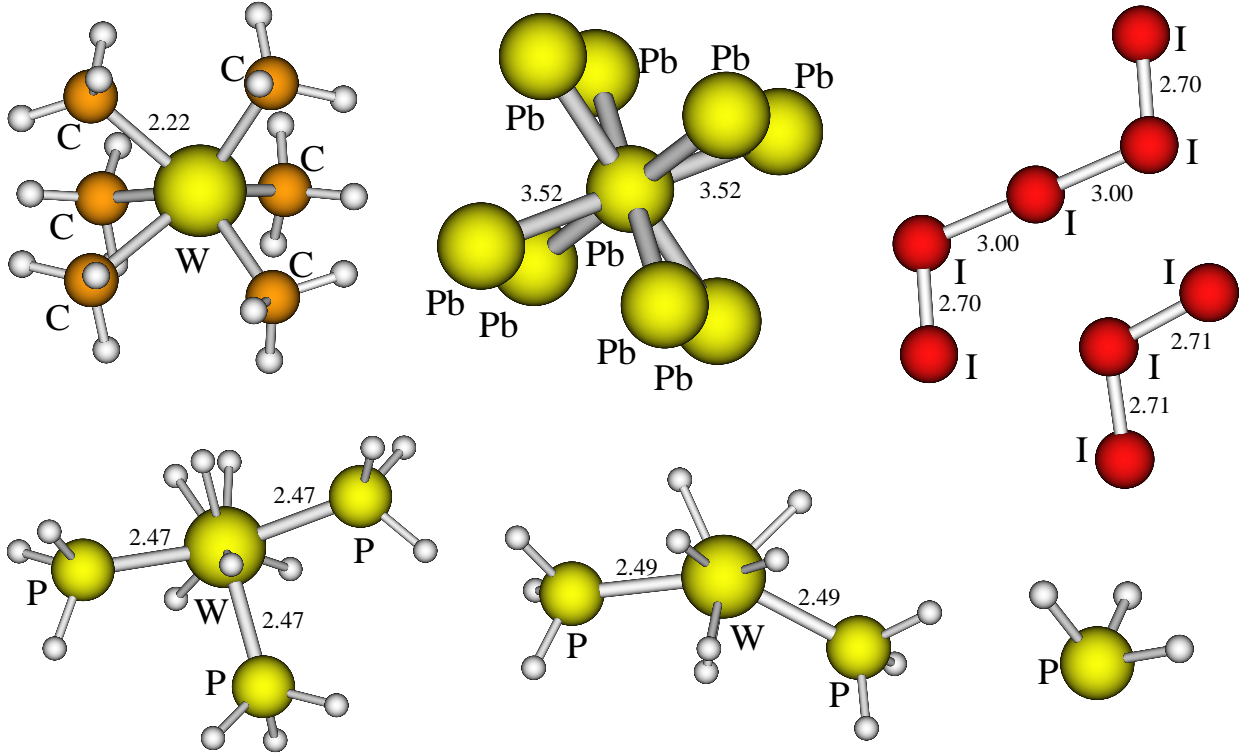}}
\caption{Structures of the test molecule set (white spheres are hydrogen atoms; selected bond lengths are given in \AA).\label{structures}}
\end{figure}

For our test molecule set, we should rely on closed-shell molecules with heavy elements in the vicinity of other
heavy elements and of light elements. Four molecules were selected to test the validity of the different local schemes:
I$_{5}^{+}$, WH$_{6}$(PH$_{3}$)$_{3}$, W(CH$_3$)$_6$ and Pb$_9^{2+}$.
Moreover, two reactions
\begin{eqnarray}
{\rm I}_5^+ &\longrightarrow & {\rm I}_3^{+} + {\rm I}_2 \\
{\rm WH}_6({\rm PH}_{3})_3 &\longrightarrow & {\rm WH}_6({\rm PH}_{3})_2 + {\rm PH}_{3}
\end{eqnarray}
were chosen to study reaction energies.
Their structures (see Fig.\ \ref{structures}) have been optimized with the {\sc Turbomole} program package \cite{Ahlrichs1989}
employing the BP86 density functional \cite{Becke1988,Perdew1986} with a valence triple-zeta basis set with polarization functions on all
atoms \cite{Schaefer1994} in combination with Stuttgart effective core potentials for 
W, Pb, and I as implemented in {\sc Turbomole}.

All molecules are closed-shell and for our analysis {\it all-electron} Hartree--Fock
calculations were then performed with the {\sc Molcas} programme package.
Since our aim is to test the validity of local relativistic approximations
instead of obtaining accurate results which can be compared to experiments,
we do not need to consider electron-correlation effects in our calculations.
The all-electron atomic natural orbital (ANO) basis sets \cite{roos05_jpca,roos05_cpl,roos08_jpca}
were employed at a double-zeta level for all atoms.
The exact-decoupling DKH calculations were performed at 35-th order,
since higher order results did not improve the accuracy for our purpose
so that this high DKH order can be considered exact.
Results of finite-order (standard) DKH2 calculations are also given for the sake of comparison.

\begin{table}[http]
\caption{Computation times (in seconds) for the relativistic one-electron Hamiltonian in various local schemes (for Pb$_9^{2+}$).} 
\label{tab_cpu}
\begin{ruledtabular}
\begin{tabular}{lrrrr}
 &
\multicolumn{1}{c}{DKH35} &
\multicolumn{1}{c}{BSS} &
\multicolumn{1}{c}{X2C} &
\multicolumn{1}{c}{DKH2}  \\
\hline
DLH(ALL)    &   1216  &   6   &   6  &  1    \\
DLU(ALL)    &   1225  &  15   &  15  & 11    \\
FULL   & 124900  & 564   & 520  & 92    \\ 
\end{tabular}
\end{ruledtabular}
\end{table}

The computation times on the Intel i7-870 CPU for the different local schemes are compared at the example of Pb$_9^{2+}$.
Table \ref{tab_cpu} presents the timings for the evaluation of the relativistic one-electron Hamiltonian matrix operator.
(the picture change transformation of property integrals are not considered in the measurement of the CPU time).
For comparison, the computation times for the one-electron and for the electron repulsion
integrals are 8s and 29744s, respectively. They are, of course, the same for the different relativistic approaches.
Since the relativistic transformations are performed on each atomic block in the DLH scheme,
its computational costs are much smaller than the FULL scheme.
The ratio of DLH to FULL is almost a constant (around 1\% in this case) for all relativistic approaches.
The DLU scheme has an extra transformation step for off-diagonal blocks in addition to the DLH scheme,
it therefore requires a constant time (about 10 second in this case) more than the DLH scheme,
which is still negligible when compared to the FULL scheme.
The computation times for the evaluation of both one-electron integrals and relativistic transformations (except for DKH35)
are smaller than that for two-electron integrals.
The relativistic transformation is usually not the bottleneck step.
However, if density functional theory calculations --- with a pure generalized-gradient-approximation functional in  
combination with some acceleration scheme like density fitting or fast multipole approximations ---  instead of
Hartree--Fock calculations were performed,
the computation time of the relativistic transformation would be significant
and the local approximation would then be required. Of course, this also holds especially for much larger molecules with light atoms (see also
discussion below).
For the high-order DKH method, like DKH35 employed here,     
the computational cost of the FULL scheme even exceed that of the two-electron integrals
and therefore a local approximation is mandatory. 

We performed DLU(ALL) and DLH(ALL) calculations
for all four molecules in order to analyze the general error that is
introduced by these local approximations. The
differences of total electronic energies with respect to
the FULL reference calculation without any local approximation are presented in Table \ref{tab_totalene}.
Since the errors of total energies are rather small,
all values are given in milli-Hartree (mH) atomic units.
The errors in total energies can be understood as 
errors on atomization energies that one would observe when atomization energies 
are to be calculated with one of the local approaches compared to the FULL one.

\begin{table}[H]
\caption{Total energy differences introduced by local approximations DLU(ALL) and DLH(ALL) with
respect to the FULL reference energy. All values are in milli-Hartree atomic units.}
\label{tab_totalene}
\begin{ruledtabular}
\begin{tabular}{llrrrr}
Method & Approximation &
\multicolumn{1}{c}{I$_{5}^{+}$} &
\multicolumn{1}{c}{WH$_{6}$(PH$_{3}$)$_{3}$} &
\multicolumn{1}{c}{W(CH$_3$)$_6$} &
\multicolumn{1}{c}{Pb$_9^{2+}$}  \\
\hline
DKH35 & FULL           & -35562819.0554  & -17170991.3602  & -16375243.0202  & -187905906.7440  \\
      & DLU(ALL)--FULL & -0.0119  & -0.0126   & -0.0204  & 0.0376   \\
      & DLH(ALL)--FULL & 0.2471   & -30.8580  & -6.3823  & -2.2619  \\ \hline
BSS   & FULL           & -35562831.5438  & -17171044.3875  & -16375296.0488  & -187907231.0132  \\
      & DLU(ALL)--FULL & -0.0119  & -0.0125   & -0.0204  & 0.0387   \\
      & DLH(ALL)--FULL & 0.2461   & -30.8303  & -6.3749  & -2.2477  \\ \hline
X2C   & FULL           & -35561841.0814  & -17171053.6272  & -16375311.3620  & -187911251.8227  \\
      & DLU(ALL)--FULL & -0.0133  & -0.0306   & -0.0367  & 0.0523   \\
      & DLH(ALL)--FULL & 0.2629   & -33.3336  & -7.0463  & -2.8362  \\ \hline
DKH2  & FULL           & -35553068.6409  & -17158776.3373  & -16363031.9017  & -187712096.3589  \\
      & DLU(ALL)--FULL & -0.0083  & 0.0011    & 0.0143   & 0.0122   \\
      & DLH(ALL)--FULL & 0.2897   & -33.4347  & -6.9555  & -3.3895  \\
\end{tabular}
\end{ruledtabular}
\end{table}

For the I$_{5}^{+}$ molecule, which possesses a pseudo-one-dimensional structure with iodine
atoms connected to at most two other iodine atoms, the DLU(ALL) scheme produces an error of
about 0.012 mH for both DKH and BSS, while X2C has a sightly larger error of 0.013 mH.
Interestingly, it is smallest for DKH2, i.e., only 0.008 mH.
The DLH(ALL) local scheme produces relatively large errors on the total electronic energies,
they are 0.24 mH for DKH and BSS, 0.26 mH for X2C, and 0.29 mH for DKH2.
Clearly, DLU(ALL) provides more accurate total energies than DLH(ALL), as it
includes the relativistic form of the atom--other-atom
off-diagonal blocks, while DLH(ALL) only takes into account
the atom--same-atom diagonal blocks.
The errors of DLH(ALL) are about 20 times larger than those of DLU(ALL) for the I$_{5}^{+}$ molecule.
This indicate that the relativistic corrections to off-diagonal blocks are quite important.
Apparently, DLU(ALL) provides an accurate scheme for taking the
off-diagonal relativistic description into account.

The Pb$_9^{2+}$ molecule features a more compact structure where more than two heavy atoms
bind to a central lead atom. Moreover, lead atoms are heavier than iodine atoms and relativistic
effects are more pronounced.
As we can see from Table \ref{tab_totalene},
errors of Pb$_9^{2+}$ are all larger than those observed for the I$_{5}^{+}$ molecule.
The deviations of DLU(ALL) electronic energies from the FULL reference energy are 0.038 mH for BSS and DKH,
while X2C is still slightly larger with a value of 0.052 mH (again DKH2 features the smallest deviation of only
0.012 mH).
However, deviations of DLH(ALL) results from the reference have become significantly larger,
they are 2.27, 2.25, and 2.84 mH for DKH, BSS, and X2C, respectively.
Hence, the errors of DLU(ALL) are roughly 60 times smaller than those observed for DLH(ALL).

Considering now molecules that also contain light atoms, the relative accuracy of DLU(ALL) vs.\ DLH(ALL)
changes dramatically. First of all, we note that our DLU(ALL) scheme preserves its accuracy 
and the deviations from the FULL reference electronic energy are in between those obtained for 
I$_{5}^{+}$ and Pb$_9^{2+}$ with the same trends as already discussed.
The errors of DLH(ALL) energies obtained for the WH$_{6}$(PH$_{3}$)$_{3}$ molecule
are, however, larger than 30 mH. I.e., they are significantly larger than those observed for Pb$_9^{2+}$.
For W(CH$_3$)$_6$, the deviations from the reference are around 6 mH, which are also still larger than for Pb$_9^{2+}$
and certainly nonnegligible.
Note that the DLH(ALL) errors obtained for WH$_{6}$(PH$_{3}$)$_{3}$ and W(CH$_3$)$_6$
are not all below those of the Pb$_9^{2+}$ molecule
although the nuclear charge number of tungsten is smaller than that of lead. This is in
contrast to the results obtained for our DLU(ALL) scheme.
Since DLH neglects any relativistic correction to off-diagonal 'interatomic' blocks,
the large errors must be due to the strong coupling between tungsten and its
surrounding lighter atoms (phosphorus and carbon).
Apparently, if a heavy-element-containing molecule has several strong bonds to this heavy element,
the off-diagonal blocks of neighboring atom pairs can have significant contributions
to the total electronic energy.
Therefore, the DLH scheme may then introduce non-negligible errors.

Since DLU(ALL) gives a balanced description for all blocks,
it will still work in those cases where DLH(ALL) fails.
DLU can therefore even be recommended for calculations of molecular systems in different geometries,
as for example required for the calculation of potential energy surfaces.
DLH would be unfit for this case as it will introduce large errors
when the inter-atomic distances become short.

DLU(ALL) also works well for the finite-order (standard) DKH2 method.
The errors are even smaller compared to exact-decoupling approaches.
However, note that the reference energies of the different relativistic approaches are not the same.
Since the DLU(ALL) scheme only discards the off-diagonal blocks of the
decoupling transformation compare to the FULL reference scheme,
the small errors in the total electronic energies indicate that neglecting off-diagonal terms
in the decoupling transformation produces negligible errors for the total energies.
This conclusion holds for both exact-decoupling and finite-order DKH approaches.
With the DLH(ALL) scheme, however, the errors in the DKH2 calculation that are due to the
DLH approximation are of the same order as those introduced by the finite, i.e., second order
when compared to the exact-decoupling approaches.
This observation leads to the very important conclusion that none of the exact-decoupling approaches is superior
to a finite-order DKH approach if one employs a DLH scheme as this would render the accuracy gained
by exact decoupling meaningless. In fact, if DLH is for some reason the local approximation of
choice, a standard DKH2 approach would yield a more efficient relativistic scheme and nothing would
be gained by achieving exact decoupling.

Besides total energies, we also study the effects of
local approximations on molecular properties.
We calculated the picture change corrected quadrupole moments.
The property matrices are transformed according to Eq.\ (\ref{proptrans}).
The values for the components of the quadrupole moment depend on the orientation of the molecule.
Therefore, we present the spherically averaged quadrupole moments
defined as $\sqrt{(Q_{\rm XX}^2+Q_{\rm YY}^2+Q_{\rm ZZ}^2)/3}$,
in Table \ref{tab_isoquad}.
Where $Q_{\rm XX}$ denotes the $XX$ (diagonal) component of the quadrupole moment.
The spherically averaged quadrupole moments are isotropic
and thus independent of molecular orientation.

\begin{table}[H]
\caption{Spherically averaged quadrupole moments for our test set of molecules.
 All values are given in Hartree atomic units.}
\label{tab_isoquad}
\begin{ruledtabular}
\begin{tabular}{llrrrr}
Method & Approximation &
\multicolumn{1}{c}{I$_{5}^{+}$} &
\multicolumn{1}{c}{WH$_{6}$(PH$_{3}$)$_{3}$} &
\multicolumn{1}{c}{W(CH$_3$)$_6$} &
\multicolumn{1}{c}{Pb$_9^{2+}$}  \\
\hline
DKH35& FULL           & 83.84331  & 57.61285  & 57.29508  & 173.68894  \\
     & DLU(ALL)--FULL & -0.00009  & 0.00018   & 0.00025   & 0.00416    \\
     & DLH(ALL)--FULL & 0.02176   & -0.02448  & -0.00218  & -0.20252   \\ \hline
BSS  & FULL           & 83.84329  & 57.61285  & 57.29507  & 173.68801  \\
     & DLU(ALL)--FULL & -0.00009  & 0.00018   & 0.00025   & 0.00414    \\
     & DLH(ALL)--FULL & 0.02177   & -0.02449  & -0.00218  & -0.20248   \\ \hline
X2C  & FULL           & 83.84406  & 57.61240  & 57.29499  & 173.68292  \\
     & DLU(ALL)--FULL & -0.00017  & 0.00029   & 0.00027   & 0.00590    \\
     & DLH(ALL)--FULL & 0.02199   & -0.02483  & -0.00197  & -0.21653   \\ \hline
DKH2 & FULL           & 83.85310  & 57.61393  & 57.29674  & 173.80677  \\
     & DLU(ALL)--FULL & 0.00009   & -0.00012  & 0.00018   & -0.00003   \\
     & DLH(ALL)--FULL & 0.02196   & -0.02458  & -0.00186  & -0.20760   \\
\end{tabular}
\end{ruledtabular}
\end{table}

The relative errors of DLU(ALL) are all below 0.001\%,
this upper limit can be further reduced to 0.0001\%
if we do not include Pb$_9^{2+}$.
Hence, DLU(ALL) is not only a good approximation for total electronic energies
but also for molecular properties like quadrupole moments.
The errors introduced by DLU(ALL) are all negligible,
whereas DLH(ALL) shows large errors for quadrupole moments.
While for total electronic energies, the largest error of the DLH(ALL) scheme shows up
in WH$_{6}$(PH$_{3}$)$_{3}$ and the smallest in I$_{5}^{+}$, it is, for the quadrupole moments,
largest for Pb$_9^{2+}$ and smallest for W(CH$_3$)$_6$.
This different behavior of DLH(ALL) indicates again the important contribution of off-diagonal
'interatomic' blocks, whose importance must be different for electronic energies and for properties.
In our DLU(ALL) scheme, I$_{5}^{+}$ has the smallest error and
Pb$_9^{2+}$ the largest one for both, energies and properties.
Therefore, DLU(ALL) turns out to give a consistent description of relativistic effects,
while DLH(ALL) yields an unbalanced relativistic treatment.
We should stress that we have also investigated other properties (radial moments, electric field gradients, contact densities) 
drawing basically the same conclusions (which is the reason, why we do not report those data here).

For both, energies and quadrupole moments, in the DLU(ALL) scheme,
DKH and BSS yields almost the same deviations from the reference
and X2C always has larger ones than DKH and BSS.
This similarity of DKH and BSS can be well understood
since both of them employ the free-particle FW transformation
as well as the subsequent exact-decoupling transformation.
The close results of DKH and BSS were also observed in Ref.\ \cite{peng12_tca}.
The slightly larger errors of X2C indicate that the off-diagonal blocks
of the X2C exact-decoupling transformation have slightly larger contributions
compared to DKH and BSS.
However, such differences of errors are too small to be considered a serious 
drawback for the X2C approach;
they are smaller than the differences of the total values.
We may use WH$_{6}$(PH$_{3}$)$_{3}$ as an example.
The difference of total isotropic quadrupole moments 
(57.61285$-$57.61240) is 0.45 milli atomic units,
while the difference of errors by DLU(ALL) 
(0.00029$-$0.00018) is 0.11 milli atomic units.
The errors introduced by the DLU(ALL) approximation are smaller than
the differences between exact-decoupling approaches.
This is also the case for the differences between DKH2 and
exact-decoupling approaches.
As we can see, the difference between DKH2 and BSS
of total isotropic quadrupole moments 
(57.61393-57.61285) is 1.08 milli atomic units,
which are much larger than the DLU(ALL) errors.

For chemical purposes, comparisons of total electronic energies are meaningless,
because chemical reaction kinetics and thermodynamics are governed by energy differences.
We therefore calculated the electronic reaction energies of the two reactions mentioned above.
The results are given in Table \ref{tab_rei5} and Table \ref{tab_rewp3}.
The errors of relative energies are less than those of absolute total energies.
For example with the DKH approach,
the error of I$_{5}^{+}$ is reduced from 0.012 mH to 0.002 mH for DLU(ALL),
0.24 mH to 0.12 mH for DLH(ALL).
For the WH$_{6}$(PH$_{3}$)$_{3}$ molecule,
the error is decreased from 0.0126 mH to 0.0006 mH for DLU(ALL),
30.9 mH to 19.4 mH for DLH(ALL).
This also holds for other relativistic approaches.

\begin{table}[H]
\caption{Electronic Hartree--Fock reaction energies for the reaction
${\rm I}_5^+ \longrightarrow  {\rm I}_3^{+} + {\rm I}_2$.
All values are in milli-Hartree (note that 1 mH is equivalent to about 2.6 kJ/mol).
}
\label{tab_rei5}
\begin{ruledtabular}
\begin{tabular}{lrrrr}
 &
\multicolumn{1}{c}{DKH35} &
\multicolumn{1}{c}{BSS} &
\multicolumn{1}{c}{X2C} &
\multicolumn{1}{c}{DKH2}  \\
\hline
FULL & 13.9797  & 13.9798  & 13.9739  & 13.9550   \\
DLU(ALL)--FULL & 0.0022  & 0.0022  & 0.0026  & 0.0011   \\
DLU(NB)--FULL & 0.0040  & 0.0040  & 0.0044  & 0.0027   \\
DLH(ALL)--FULL & -0.1190  & -0.1190  & -0.1214  & -0.1231   \\
\end{tabular}
\end{ruledtabular}
\end{table}

Next, we study the neighboring-atomic-block (NB) approximation at the example of the I$_{5}^{+}$ molecule.
As shown in Figure \ref{structures}, it is composed of a chain of iodine atoms
and thus suitable for applying the neighboring approximation.
Only the pairs of atoms which are connected with a bond displayed
in Figure \ref{structures} are considered as neighbors.
The neighboring groups could also be determined by
introducing a cut-off distance parameter.
Then, the pair of atoms which has shorter distance than the
given parameter is counted as a neighboring pair.
The relativistic transformations are now only applied to such neighboring pairs 
according to Eq.\ (\ref{dlunbtrans}) within the DLU scheme.
As we can see from Table \ref{tab_rei5},
the neighboring approximation DLU(NB) gives very small errors for
relative energies.
Therefore it is a very good (additional) approximation within our DLU scheme.
Hence, with DLU(NB) the computational cost will become linear scaling,
while the DLU(ALL) scheme has an order-$N^2$ scaling,
which may be a bottleneck for calculations on very large molecules.
But be aware that the success of the NB approximation depends on how
one defines the neighboring atoms.
If the cut-off distance is too small, the DLU scheme will be reduced
to the DLH scheme since no neighbors would be found.

Reduction in terms of computational costs can also be achieved by employing
the BP or NR approximation to the light atoms.
It should be a good approximation if a molecule has many light atoms
such as hydrogen.
We study these approximations for the phosphine ligand dissociation energy of
WH$_{6}$(PH$_{3}$)$_{3}$.
The results are presented in Table \ref{tab_rewp3},
where DLU(WP,NR) denotes the NR approximation applied to the H atoms
while W and P atoms are relativistically described.
By contrast, DLU(W,NR) denotes application of the relativistic transformation only to
the W atom, while for P and H the NR approximation applies.
From Table \ref{tab_rewp3} we can see that the results of the BP approximation are 
very bad,
especially if the BP approximation was employed for the phosphorus atoms.
The BP approximation to P and H atoms gives errors which are
roughly seven times the value for the reactions energy. Hence, this is approximation turns out to be
completely useless. 
Such huge errors stem from the variational collapse of the BP approximation.  

\begin{table}[H]
\caption{Relative energy difference of the reaction
${\rm WH}_6({\rm PH}_{3})_3 \longrightarrow  {\rm WH}_6({\rm PH}_{3})_2 + {\rm PH}_{3}$.
All values are in milli-Hartree.}
\label{tab_rewp3}
\begin{ruledtabular}
\begin{tabular}{lrrrr}
&
\multicolumn{1}{c}{DKH35} &
\multicolumn{1}{c}{BSS} &
\multicolumn{1}{c}{X2C} &
\multicolumn{1}{c}{DKH2}  \\
\hline
FULL & 35.8447  & 35.8448  & 35.8439  & 35.9064   \\
DLU(ALL)--FULL & -0.0006  & -0.0007  & 0.0040  & 0.0101   \\
DLU(WP,NR)--FULL & -0.1579  & -0.1579  & -0.1533  & -0.1476   \\
DLU(W,NR)--FULL & 0.3648  & 0.3648  & 0.3684  & 0.3737   \\
DLU(WP,BP)--FULL & 1.2501  & 1.2500  & 1.2547  & 1.2608   \\
DLU(W,BP)--FULL & 254.4914  & 254.4914  & 254.4951  & 254.5018   \\
DLH(ALL)--FULL & 19.4202  & 19.4024  & 21.0255  & 21.1635   \\
DLH(WP,NR)--FULL & 19.4229  & 19.4051  & 21.0284  & 21.1663   \\
DLH(W,NR)--FULL & 19.6722  & 19.6544  & 21.2789  & 21.4146   \\
DLH(WP,BP)--FULL & 20.3668  & 20.3490  & 21.9743  & 22.1116   \\
DLH(W,BP)--FULL & 273.0474  & 273.0296  & 274.6570  & 274.7922   \\
\end{tabular}
\end{ruledtabular}
\end{table}

However, the NR approximation is much better.
The errors on the reaction energy within the DLU(NR) scheme, which are in the range of 0.1 to 0.3 mH and
thus less than 1 kJ/mol, are perfectly acceptable.
This success is due to the correct nonrelativistic limit of the NR approximation within the DLU scheme.
Not unexpectedly, the DLH(NR) results have quite larger errors,
because DLH(ALL) already does not provide good results for the
phosphine dissociation energy from the WH$_{6}$(PH$_{3}$)$_{3}$ complex.
If we study the difference between DLH(NR) and DLH(ALL),
we find that the NR approximation to actually give small errors
on top of DLH(ALL).
However, the DLH scheme is not recommended since its accuracy is not guaranteed at all.
Only in those cases for which the DLH scheme may work,
we can further add the NR approximation on top of it for light atoms.
As can be seen from both Table \ref{tab_rei5} and Table \ref{tab_rewp3},
different relativistic approaches behave basically identical with respect to the discussion of
the NB and NR approximations.
Therefore, the approximations for the reduction of computational cost
discussed in this article can be applied to all relativistic transformation approaches.

\section{Conclusions}
In this work, we aimed at a rigorous local approximation to relativistic transformation schemes,
which makes them applicable to and efficient for calculations on large molecules. We developed
a systematic hierarchy of approximations that is based on the assembly of the unitary transformation
from 'atomic' contributions (DLU) rather than on a local approximation directly applied to the matrix
representation of the Hamiltonian (DLH).

The straightforward DLH scheme, which has evolved in the development of the DKH method (see references given above),
turns out to be not a very accurate local approximation since it only covers the relativistic treatment of the atom--same-atom
diagonal blocks in the Hamiltonian. It may fail if inter-atomic distances become short so that
a relativistic treatment of the off-diagonal 'interatomic' blocks becomes important.
As a consequence, the DLH approximation is not suitable, for example, in studies
that crucially depend on accurate electronic energy surfaces.
Even if the molecular geometry is fixed,
the DLH approach does not provide a balanced description for different molecular properties,
since the contributions of the off-diagonal blocks to total observables
are quite different for different properties.

By contrast, the DLU scheme proposed by us in this paper
overcomes the drawbacks of the DLH scheme.
It is an excellent approach to take into account the atom--other-atom (i.e., 'interatomic')
off-diagonal relativistic transformations.
It does not show an instability for varying
molecular structures (i.e, for electronic energy surfaces) and properties.
The size of the error introduced by DLU is much smaller than that of DLH;
it is typically only 1\% of the latter.
The errors of the DLU approach when compared to the full relativistic transformation
without any local approximation turn out to be tiny.
They are even smaller than the difference among the different exact-decoupling approaches,
which are already very small.

The DLU scheme is valid for all exact-decoupling approaches,
while the local $X$-matrix scheme only works for the X2C approach.
Furthermore, the DLU scheme can also be applied to finite-order
DKH approaches such as DKH2,
which has been very successful in computational chemistry and whose computational costs are less
than those of any exact-decoupling approach.

If one has to use the DLH scheme for some reason, such as the
lack of a DLU implementation or because of its linear-scaling behavior,
the selection of a relativistic approaches is not decisive as
the errors introduced by the DLH scheme are already higher than the difference
of different relativistic approaches. Hence, the approximate, but fast DKH2 method
may be safely combined with the DLH approximation, while this cannot be recommended for any
exact-decoupling approach.

However, the linear-scaling behavior of DLH is no good reason 
since the DLU scheme can also be made linear-scaling within the 
neighboring-atomic-blocks approximation, which produces
negligible errors.
One can further reduce the computational costs by
employing the BP or NR approximation to all light atoms.
However, the BP approximation turns out to be not suitable
for variational calculations since it introduces large errors.
Only the NR approximation yields acceptable results.
The errors of the NR approximation will be larger
if the nuclear charge of the atom for which the NR approximation is invoked increases.
Therefore, the application of the NR approximation will depend on the
balance of accuracy and computational cost in an actual calculation.

While this work focused on the accurate approximation of exact-decoupling methods at the self-consistent-field level
with respect to total electronic energies and first-order properties of a small, but well-selected test molecule set,
further investigations into the DLU scheme concerning, for example, higher-order properties, spin--orbit and
electron-correlation effects are under way in our laboratory.

\begin{acknowledgements}
This work has been financially supported by the Swiss national science foundation.
\end{acknowledgements}


\end{document}